\documentclass{aa}  

\usepackage{graphicx}
\usepackage{txfonts}
\usepackage{soul}
\usepackage{xcolor}
\usepackage[breaklinks=true]{hyperref} 
\hypersetup{final,colorlinks=true,linkcolor=blue,citecolor=blue,urlcolor=blue} 

\makeatletter
\renewcommand*\aa@pageof{, page \thepage{} of \pageref*{LastPage}}
\makeatother

\begin{document} 

\title{Solar Jet Hunter: a citizen science initiative to identify coronal jets in EUV data sets}

\titlerunning{Solar Jet Hunter: a citizen science initiative to identify coronal jets}

\author{S. Musset \inst{1,}
  \and P. Jol \inst{2} 
  \and R. Sankar \inst{3,4}
  \and S. Alnahari \inst{3,4}
  \and C. Kapsiak \inst{3}
  \and E. Ostlund \inst{3,4}
  \and K. Lasko \inst{3,4}
  \and L. Glesener \inst{3,4}
  \and L. Fortson \inst{3,4}
  \and G. D. Fleishman \inst{5,6}
  \and Navdeep K. Panesar \inst{7,8}
  \and Y. Zhang \inst{3,4} 
  \and M. Jeunon \inst{9}
  \and N. Hurlburt \inst{7}}

\institute{European Space Agency, European Space Research and Technology Centre, Keplerlaan 1, 2201 AZ, Noordwijk, The Netherlands
    \and Leiden Observatory, Leiden University, PO Box 9513, 2300 RA, Leiden, the Netherlands
    \and School of Physics and Astronomy, University of Minnesota, 116 Church St. SE, Minneapolis, MN 55455, USA
    \and Minnesota Institute for Astrophysics, University of Minnesota, 116 Church St. SE, Minneapolis, MN 55455, USA
    \and New Jersey Institute of Technology, University Heights, Newark, NJ 07102-1982, USA
    \and Leibniz-Institut für Sonnenphysik (KIS), Freiburg, 79104, Germany
    \and Lockheed Martin Solar and Astrophysics Laboratory, 3251 Hanover Street Building 252, Palo Alto, CA 94304, USA
    \and Bay Area Environmental Research Institute, NASA Research Park, Moffett Field, CA 94035, USA
   \and The Catholic University of America, 620 Michigan Avenue, Washington, DC 20064, USA}

   \date{Received , ; accepted , }


\abstract
{Solar coronal jets seen in EUV are ubiquitous on the Sun, have been found in and at the edges of active regions, at the boundaries of coronal holes, and in the quiet Sun. Jets have various shapes, sizes, brightness, velocities and duration in time, which complicates their detection by automated algorithms. So far, solar jets reported in the Heliophysics Event Knowledgebase (HEK) have been mostly reported by humans looking for them in the data, with different levels of precision regarding their timing and positions.}
{We create a catalogue of solar jets observed in EUV at 304 \AA \  containing precise and consistent information on the jet timing, position and extent.}
{We designed a citizen science project, "Solar Jet Hunter", on the Zooniverse platform, to analyze EUV observations at 304~\AA \  from the Solar Dynamic Observatory/Atmospheric Imaging Assembly (SDO/AIA). We created movie strips for regions of the Sun in which jets have been reported in HEK and ask the volunteers to 1) confirm the presence of at least one jet in the data and 2) report the timing, position and extent of the jet.}
{We report here the design of the project and the results obtained after the analysis of data from 2011 to 2016. 365 "coronal jet" events from HEK served as input for the citizen science project, equivalent to more than 120,000 images distributed into 9,689 "movie strips". Classification by the citizen scientists resulted with only 21\% of the data containing a jet, and 883 individual jets being identified. }
{We demonstrate how citizen science can enhance the analysis of solar data with the example of Solar Jet Hunter. The catalogue of jets thus created is publicly available and will enable statistical studies of jets and related phenomena. This catalogue will also be used as a training set for machines to learn to recognize jets in further data sets.}


\maketitle
\section{Introduction}

The Sun fills the heliosphere with a continuous flux of plasma, the solar wind, and with energetic particles, accelerated during eruptive events such as solar flares and coronal mass ejections. In the case of solar flares, energetic particles are accelerated in the low corona, and the means by which they are injected into the interplanetary medium are still a topic of investigation. 
Coronal jets are collimated ejections of plasma in the solar atmosphere, often detected in  X-rays \citep[e.g.][]{alexander_1999,moore_2018},  extreme ultraviolet (EUV; \citealt{nistico_2009,panesar_2016a,sterling_2016}), and radio \citep{2018ApJ...867...84G, Kaltman_etal_2021}. These ejections illuminate magnetic field lines open towards the interplanetary medium, which offer a path for accelerated charged particles to escape the solar atmosphere and reach the interplanetary medium. Therefore, jets are thought to be good tracers of the injection of flare-accelerated particles in the heliosphere. A few studies of solar energetic particle (SEP) events detected in-situ have indeed found jets associated with the flares or CMEs thought to be at the origin of these SEPs \citep{Krucker_etal_1999, Dresing_etal_2021, Nitta_etal_2015, Bucik_2022}.
Another indication of the association between coronal jets and the escape of energetic particles from the solar corona is the correlation found between jets and type III radio bursts, signatures of escaping beams of accelerated electrons in the high corona and interplanetary medium. \cite{Christe_etal_2008} showed that jets were closely related in time to type III radio bursts, and \cite{Glesener_etal_2012} used the imaging capabilities of the Nançay Radioheliograph (NRH) to demonstrate in one case that the type III emission was emitted on open field lines associated with a jet seen in EUV. In addition to being sources of energetic particles for the heliosphere, jets may play a significant role in driving the solar wind \citep{raouafi2023} in the form of small-scale reconnection episodes along field lines open to interplanetary space.

Coronal jets are ubiquitous in the corona, having been detected in and around the edges of active regions \citep{Shibata_etal_1992,sterling_2016,odermatt_etal_2022}, at the boundaries of coronal holes \citep{panesar_2018a} and in the quiet Sun \citep{mcglasson_2019}; they have a variety of morphologies, sizes and durations \citep[see the reviews of][]{innes_2016, Raouafi_etal_2016}. These jets have been detected first in soft X-rays \citep{Shibata_etal_1992}, and are now routinely detected in EUV images at different wavelengths. Using high resolution EUV images from the Atmospheric Imaging Assembly (AIA; \citealt{lemen_12}) onboard the Solar Dynamics Observatory, many authors \citep[][]{adams_2014,Sterling_etal_2015,panesar_2016b,mcglasson_2019} reported that coronal jets are driven by the eruption of a small-scale filament (known as \textit{minifilament}), which often forms and erupts as a result of surface magnetic flux cancellation \citep{shen_2012,panesar_2016b,panesar_2017}. It has been observed that when a minifilament erupts, a brightening (known as jet base brightening) appears underneath the erupting minifilament. The jet base brightening is interpreted as a miniature version of solar flare arcade. While the jet base brightening is at the base of an erupting minifilament, a flare arcade appears in the wake of typical CME-producing filaments \citep{Sterling_etal_2015}. Therefore coronal jets are often interpreted as small-scale analogs of larger-scale CME-producing eruptions.

Many of these investigations focused on small samples of events, sometimes even just one event, and had to draw conclusions from that.  Due to these small samples, many factors remain unclear - e.g., how often solar jets are associated with the escape of particles, what role they play in the acceleration and propagation of these energetic particles, the total energy that they carry in aggregate into the heliosphere, and their influence on the energetics of the solar wind.  The lack of broad-ranging statistical studies is due to the lack of a precise, expansive, and systematically developed database of solar jets.  This is challenging because jets are not always bright, and their variety in shapes, sizes, and temperatures make their detection by automated algorithms challenging. In the Heliophysics Events Knowledgebase (HEK\footnote{\url{https://lmsal.com/hek/}}), coronal jets have been reported by hand by a relatively small number of observers.  While this is useful information (and has driven work such as \citet{musset2020}), the reports of jets identified in this way have different precision on the timing and position of the jets, and the sensitivity of detection is difficult to determine. Some "coronal jet" entries in the HEK cover several hours and are essentially a succession of jets, without precise timing information. Moreover, some prominent jets are missing from the HEK database, demonstrating that this jet database is incomplete.  The reasons for these deficiencies are the immense amount of researcher time it takes to page through data and identify jets, and the varying observational choices that each individual researcher uses as they conduct their work (choice of wavelength, field of view, time cadence, etc. to use in constructing their movies).

In order to build a consistent and complete database of jets, we require more sophisticated machine models which can both identify coronal jets in observations, and provide an approximate location and boundary of the jet. Deep learning presents a promising avenue for this effort, but these models usually require tens of thousands of samples for training data, which is hard to generate by hand by a single research team. A promising pathway to generate this training dataset is using citizen science, where members of the public (generally with little or no a-priori scientific background) can label or annotate scientific datasets, which can then be used by research teams. One of the earliest examples of such projects dates back to 1716, when volunteers across central England reported observations of the 1715 total solar eclipse \citep{Halley1716}. More recently, citizen science has been used extensively across the scientific domains. For example, in astronomy, there are several projects that are focused on data gathering, e.g., amateur observations of planets \citep{Hueso2010} or impact flashes \citep{Sankar2020}, and in classifications or annotation, e.g., classification of galaxy morphologies \citep{Lintott2008}. For a broad review of citizen science efforts in astronomy and space science, see \citet{Marshall2015} and \citet{fortson2021green}. The benefits of citizen science in classification tasks is two-fold: first, citizen scientists can be quickly trained on a small sample of pre-selected data, which is generally difficult to do with machine models. Secondly, classifications from multiple volunteers can be used for the same object in order to build a consensus value and quantify the error in the classification \citep{Lintott2008}. Citizen science has also been vital in the engagement of the community in scientific research, leading to increased science learning \citep{Masters2016} and serendipitous discoveries -- objects in the dataset that contain previously unknown scientific phenomenon \citep[see][and references therein]{Trouille2019}, which lead to interesting new breakthroughs. Ultimately, for this project, citizen science has been identified as the best method to compile a consistent catalogue of jets, which will be used both to explore the physical processes at play and associated with jets, and as a training set for machine learning on the detection of solar jets in EUV data.

Zooniverse.org is the largest citizen science platform, with over 2.5 million volunteers who provide classifications on nearly a hundred active projects, beginning with Galaxy Zoo \citep{Fortson2012}. These projects range across many disciplines from astrophysics to humanities. Zooniverse features a simple classification interface where a "subject" (image(s), video or audio) is shown to the citizen scientists and they are provided with tools to classify and/or annotate the subject. There is a library of existing tools for research teams to choose from, ranging from simple question/answer to freeform drawing, which can be used to build a classification "workflow" for the volunteers to follow. Zooniverse also features an aggregator back-end that can help automatically build consensus on the volunteer classifications. 

We created the "Solar Jet Hunter" citizen science project using ' the Zooniverse framework, with the goal of creating a consistent database of jets identified in EUV data from the Solar Dynamic Observatory (SDO). In this paper, we present how this project was setup (section \ref{sec:project-setup}), how the resulting classifications from the volunteers were processed and aggregated into a jet catalogue (section \ref{sec:aggregation}) and a preliminary analysis of the catalogue thus created (section \ref{sec:results}).

\begin{figure*}
\includegraphics[width=0.99\linewidth]{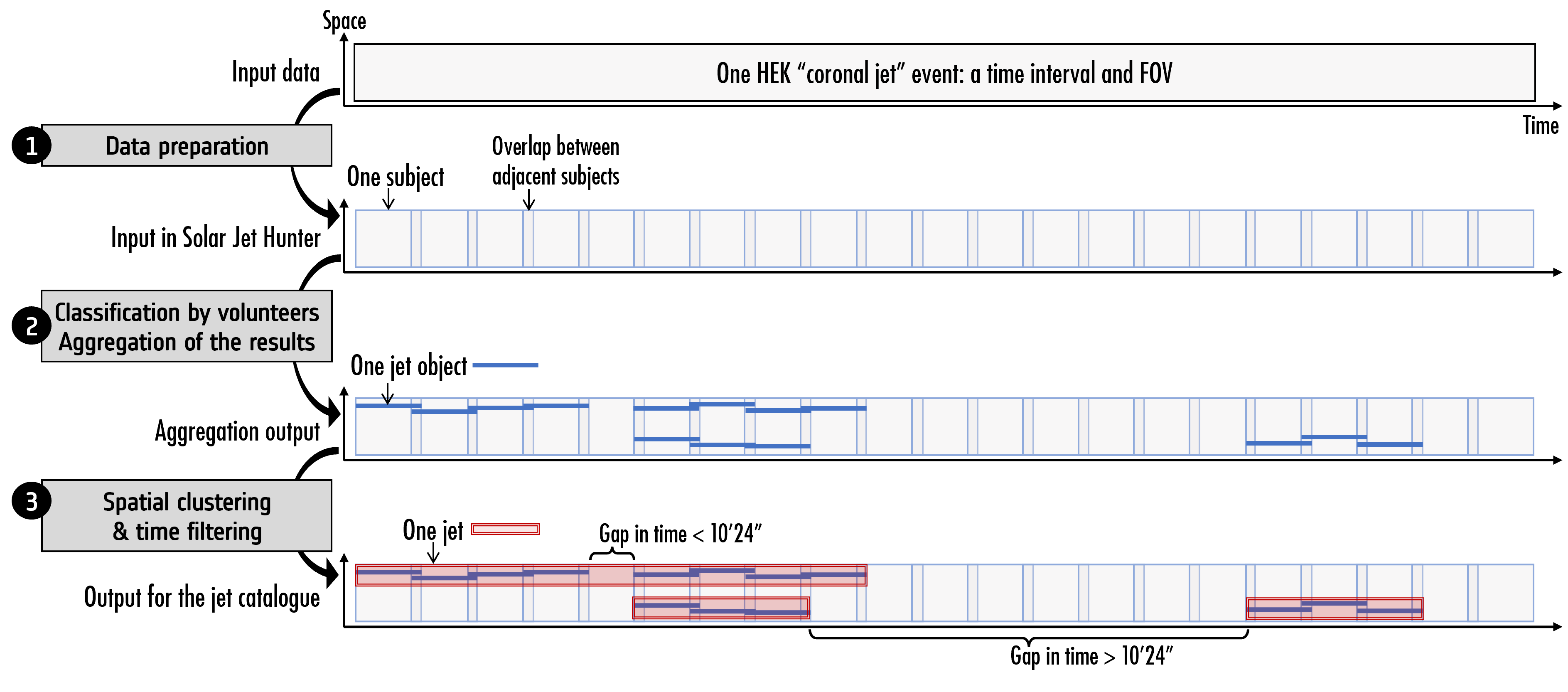}
\caption{Diagram describing the relation between HEK events, Zooniverse subjects, jet clusters and jets. The input data is a HEK event that may contain one or several jets. The time interval of the HEK event is cut into smaller time intervals to form Zooniverse subjects (see section \ref{sec:data_selection} for more details). There is an overlap of two frames between two adjacent subjects. After classification and aggregation of the results from the Solar Jet Hunter (as described in section \ref{sec:aggregaion-volunteers}), for each subject, we have a list of jet objects discovered by the volunteers. Many subjects do not contain any jet object, some subjects contain more than one jet object. Then, for all the subjects of the original HEK event, the jet objects are clustered in space and time (as described in section \ref{sec:subject_to_jets}).
In this diagram, the output is three jets, two of them overlapping in time, and two of them happening one after another at the same position.}
\label{fig:data}
\end{figure*}

\section{Zooniverse project setup}
\label{sec:project-setup}

\subsection{Data selection}
\label{sec:data_selection}

Since solar jets can be relatively small, the data we selected for the Solar Jet Hunter project are not full-Sun images. We selected times and fields of view corresponding to coronal jets reported in the HEK database. The first beta test revealed that a field of view that is \textit{too} small can make the images appear pixelated and hard to use for the volunteers, so we imposed a minimum field of view of 120 $\times$ 120 arcsec.
Moreover, jets can only be easily identified when looking at a time sequence of images, or movies: their primary identifying feature is that they are ejections of plasma. In Solar Jet Hunter, movies shown to the volunteers were limited to 15 frames (see section \ref{sec:subjects}). To ensure that the volunteers would see a long enough time interval to identify motion, we sampled the AIA data at a 24-second cadence instead of the full 12-second cadence: a movie (subject) in the Zooniverse project thus represents 6 minutes of data. Finally, we restricted our data to one AIA filter and chose the 304 \AA \ filter in which the presence of chromospheric temperature plasma makes jets easily identifiable.

With these specifications, we developed a pipeline\footnote{\tiny \url{https://github.com/kapsiak/Solar_Zooniverse_Processor}} to produce a database of subjects to be uploaded to the Solar Jet Hunter project. This pipeline performs the following steps:
\begin{enumerate}
    \item Search for the "coronal jet" instances in the HEK in a given time interval and save the results.
    \item For each HEK event, request AIA cutouts in the 304 \AA \ filter for the time interval and field of view of the event.
    \item Download the corresponding FITS files from the cutout service.
    \item Create PNG images from the FITS files, and the associated meta-data using the FITS files and HEK event properties.
\end{enumerate}

\begin{table}[]
\begin{tabular}{l|cccccc|}
\hline
\multicolumn{1}{|l|}{Year}       & 2011    & 2012    & 2013   & 2014         & 2015        & 2016  \\ \hline
\multicolumn{1}{|l|}{HEK events} & 37      & 38      & 45     & 57           & 79          & 108   \\
\multicolumn{1}{|l|}{Subjects}   & 1819    & 1299    & 819    & 988          & 1909        & 2831  \\ \hline
                                 & \multicolumn{3}{c|}{Run 1} & \multicolumn{2}{c|}{Run 2} & Run 3 \\ \cline{2-7} 
\end{tabular}
\caption{Number of HEK coronal jet events considered for each year, and corresponding number of subjects generated for the Solar Jet Hunter project.}
\label{tab:data}
\end{table}

\begin{figure*}
\includegraphics[width=0.98\linewidth]{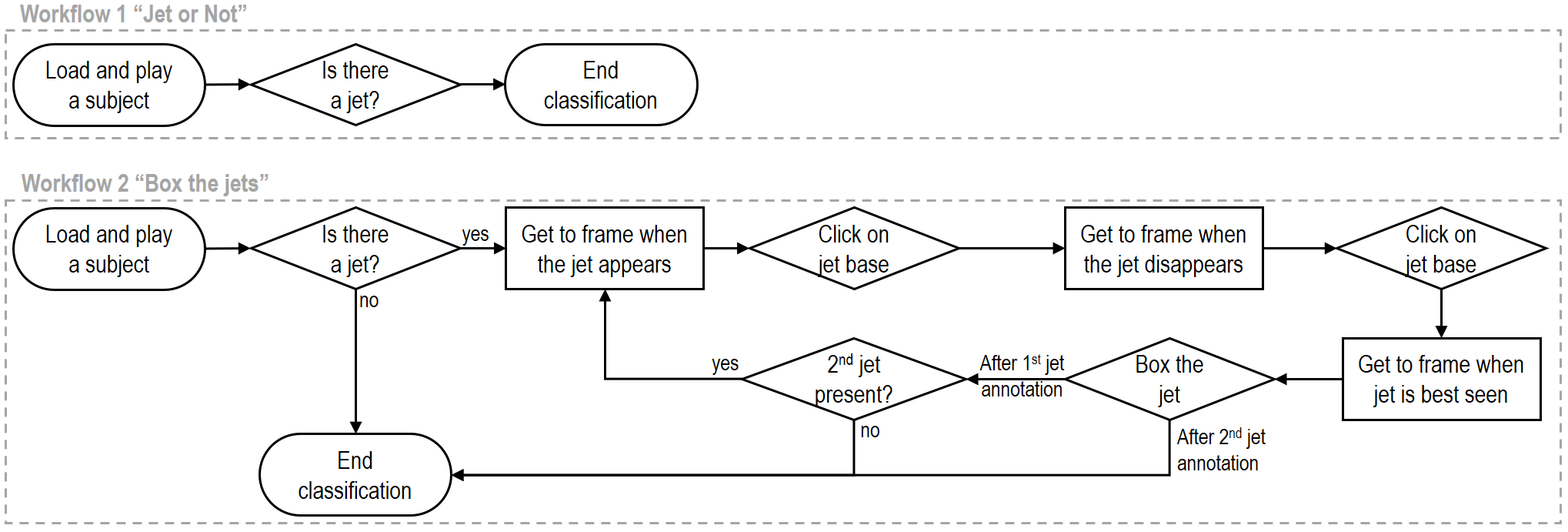}
\caption{Chart of the two parallel workflows of Solar Jet Hunter}
\label{fig:workflow}
\end{figure*}

\subsection{Solar Jet Hunter subjects}
\label{sec:subjects}

In Zooniverse, a \emph{subject} is a piece of data that will be shown to the volunteers in one or several workflows. For the Solar Jet Hunter, we present subjects as movies to the volunteers so that they can see the dynamics in the corona and spot solar jets as they evolve. It is also necessary to have tools to annotate the movie (click on the jet base for instance). When this project was designed, the Zooniverse did not have annotation tools for their video subjects. We therefore used the "movie strip" subject, which is a sequence of images that can be played as a short movie, by the volunteers. In a movie strip, the volunteer can select each frame/image individually. We chose to have 15 frames per movie strip and this number was validated as comfortable to see jets by the volunteers during our beta-test. As the project evolved, we developed a new video tool that will be used for future subject sets.

As described in section \ref{sec:data_selection}, we used AIA data with a 24-second cadence; thus a subject represents 6 minutes of data. However, solar jets often last longer than 6 minutes \citep[see e.g.][]{mulay_etal_2016, musset2020}, and the HEK events can last for hours. Therefore, each HEK event was divided into several subjects, as illustrated in Figure \ref{fig:data} (step 1). Each subject overlaps the next one by two frames. Because these subjects are quite short, it is possible that a jet spans multiple subjects. 

Subjects were uploaded to the Zooniverse platform in the form of sets of images and the associated metadata that bonded them in groups of 15 frames. The metadata also contains the names of the FITS files used to create the images as well as characteristic lengths in the images that are needed to transform pixel coordinates to solar coordinates. They were uploaded using the Panoptes\footnote{\tiny \url{https://panoptes-python-client.readthedocs.io/en/latest/}} pipeline developed by the Zooniverse team. On Zooniverse, subjects are arranged into \emph{subject sets}. We created one subject set per year of observation. In the first run of the project, our subject sets included HEK jet events from 2011 to 2013. We re-launched the project in two other runs in the course of the year 2022, with data from 2014 to 2016. The number of HEK events and the corresponding number of subjects are summarized in table \ref{tab:data}. In this paper we present the results from the three runs of the project together. The different runs of the projects are further described in section \ref{sec:results-enagagement}.

\subsection{Solar Jet Hunter workflows}

In Solar Jet Hunter, volunteers classify jets within two workflows, which exist in parallel as long as they contain data yet to be analyzed. 

The first and most simple workflow is "Jet or Not". The goal of this workflow is to identify the subjects in which at least one jet is present. The task is schematized in Figure \ref{fig:workflow} and is simple: a subject is loaded, the volunteer has to play the movie strip and then select "yes" or "no" to answer the question: "Is there a jet in these data?". During the first run of the project, the subject was retired from the workflow when 3 different volunteers answered the question.  It was determined during the aggregation that the accuracy of the answer would likely improve with a higher number of volunteer answers, so for the other runs of the project, any subject was shown to at least 7 different volunteers in this workflow before being retired. 

When subjects are retired from the first workflow, if more than half the volunteers who classified the subject answered "yes" and found at least one jet in the data, the subject is pushed into a new subject set. In this new subject set, we only have subjects which contain at least one jet. This subject set is then used as input to the second workflow. This operation is performed by the Caesar aggregation tool developed by the Zooniverse team. 

\begin{figure}
\includegraphics[width=0.98\linewidth]{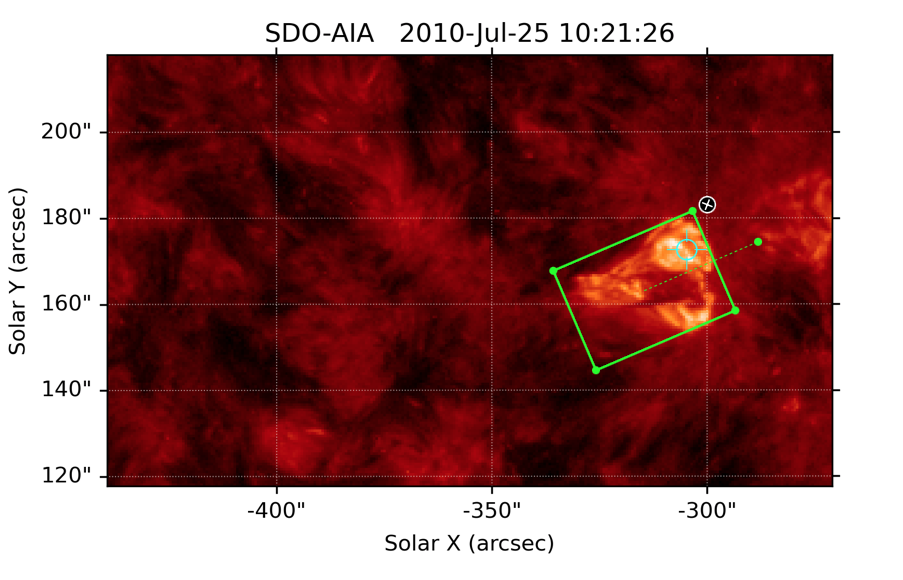}
\caption{Example of annotation in a frame of a movie strip in the workflow "Box the jets" in Solar Jet Hunter. A cyan target shows where the volunteer clicked to report the base of the jet. A green rectangle has been drawn and rotated to box the jet.}
\label{fig:workflow2}
\end{figure}

The goal of the second workflow, "Box the jets", is to analyze more closely the jets found by the volunteers in the first workflow, in order to provide position and timing information for each jet. This workflow is schematized in Figure \ref{fig:workflow}. First, we ask for a confirmation that the volunteer sees a jet in the data. Then, we ask the volunteer to report on the timing, position and extent of the jet. To achieve this, we ask the volunteer to click on the base of the jet (location) at the time when the jet first appears (timing). To do so, the volunteer has to select the first frame in which the jet is visible and then click on the base of the jet in the image. In the second task we ask the volunteer to again select the base of the jet, but at the last frame for which the jet can be seen. Finally, we ask the volunteer to choose the frame in which the jet is best seen, and to draw a rectangle around the jet, which gives us an indication of the jet extent and orientation. An example of the annotations performed by the volunteers is shown in Figure\,\ref{fig:workflow2}.
One complexity in our data is the possibility to have more than one jet in each subject. We therefore allowed for the annotation of a second jet if the volunteer answered "yes" to the question: "Did you see a second jet in the subject?". After going through the tasks for the second jet, if there is more than two jets, the volunteer is asked to flag this particular event in the talk board (see section \ref{section:vol_exp}) using the keyword \emph{\#multiplejets}. 

The "Box the jets" workflow requires a deeper involvement from the volunteers who may spend tens of seconds or even a minute on each subject, rather than the few seconds spent to classify jets in the first workflow. It also involves looking at both the temporal and spatial evolution of the jet, raising the complexity of the task. Since the classification products of the second workflow are more complex, they are more subject to noise and thus each subject is classified by at least 16 different volunteers before being retired from the workflow.

\subsection{Volunteer experience}
\label{section:vol_exp}

Citizen science relies on analysis that can be performed with minimal training, so that any volunteer can participate indifferently of their science or education background. The hunt for solar jets is a good science use case for citizen science as the main skill required is to spot moving features (ejections) in short movies showing observations of the Sun; technical knowledge of the physical principles behind the ejections is not required. However, this task is complicated by the fact that many moving features exist in these observations, and not all of them are jets. Therefore, a brief amount of training is required to enable the volunteers to identify and annotate solar jets.

The Zooniverse framework contains several features to provide training, guidance, and communication tools for the volunteers. They include pages with the description of the project goals, team and results, a tutorial, a guide field, and a talk board.

Volunteer training is provided via a tutorial that automatically pops up during the first visit of a volunteer on the Solar Jet Hunter website. This tutorial summarizes the goal of the project, gives a few examples of jets, and explains how to use the annotation tools. It also points to the other features available for the volunteers: pages that describe the project, a field guide and a talk board. The field guide is particularly important for Solar Jet Hunter as it provides additional examples of solar jets, which can have various shapes and sizes, as well as examples for many other features that can be present in the data and that are not solar jets. These features include filament eruptions, prominences, spicules, loops, flares and active regions.
The talk board is the main way for volunteers to communicate with other volunteers and with the research team behind the project. Each workflow ends with the possibility to 1) submit the answers to the questions or 2) submit and go to the talk board, to comment or ask a question regarding the specific subject currently analyzed. In Solar Jet Hunter, this is used in particular to report subjects with more than two jets.

\section{Aggregation methodology}
\label{sec:aggregation}

\subsection{Aggregation of the volunteers classifications}
\label{sec:aggregaion-volunteers}

Given that there are multiple responses by different volunteers to a given subject, there is a need to aggregate (reduce) the classifications into a single result. Zooniverse features a suite of tools for this purpose in the \texttt{panoptes\_aggregation} Python package. Here, we describe the process of reducing the volunteer responses in each workflow to obtain unique jets in each subject (step 2 in Figure \ref{fig:data}), and how jets in different subjects were clustered into jets spanning over several subjects (step 3 in Figure \ref{fig:data}). 

The `Jet or Not' workflow features a question task, requiring volunteers to provide a binary response as to whether a subject contains a jet. For each subject, we define the agreement as the fraction of users that voted for the most chosen answer, given by,
\begin{equation}
    {\rm agreement} = \dfrac{C_m}{N},
\end{equation}
\label{equation:agreement}
where $m$ is the chosen response (i.e., the one with the modal volunteer count), $C_m$ is the number of volunteers who chose $m$, and $N$ is the total number of votes. The agreement score denotes the agreement between the volunteers on the modal choice, and is a proxy for the confidence that the subject contains a jet. Agreement also denotes the inherent `difficulty' in a subject, i.e. a more difficult subject would be one where jets are much harder to distinguish against a background, making it less likely for all volunteers to spot that jet. 

For the `Box the Jets' workflow, there are several steps to the aggregation process. Volunteers are asked to annotate the base of the jet in the first and last frames in which it is visible (hereafter, the start and end bases, respectively), and also draw a bounding box around the jet when it has the largest size. This is repeated for up to two jets per subject. Volunteer annotations are then aggregated to produce sets of start and end bases and one box per jet in the subject. 
Since each subject might contain multiple jets, and the volunteers may annotate these jets in different orders, we perform the aggregation with all the annotations at once, and then disentangle individual jets after clustering. In this way, we can determine multiple jets per subject where necessary. We then track these jets across subjects to build a collection of unique jets per HEK event (see \S~\ref{sec:subject_to_jets}).

\subsubsection{Point clustering for base locations}

The volunteer annotations for the base of the jet (both start and end) are clustered using the Hierachical Density-Based Spatial Clustering of Applications with Noise algorithm \citep[HDBSCAN;][]{McInnes2017}. HDBSCAN uses the density of the points to identify clusters (using the Euclidian pixel distance between the points). Therefore, in our case, volunteer annotations of the base for the start and end of the jet are first clustered based on their proximity to each other. The aggregated value is then simply the centroid of each cluster. An example of the start and end bases are shown in Figure~\ref{fig:point_clustering}. We detail the performance of our clustering algorithm in Appendix~\ref{appendix:point_clustering}. 

\begin{figure}
    \centering
    \includegraphics[width=\columnwidth]{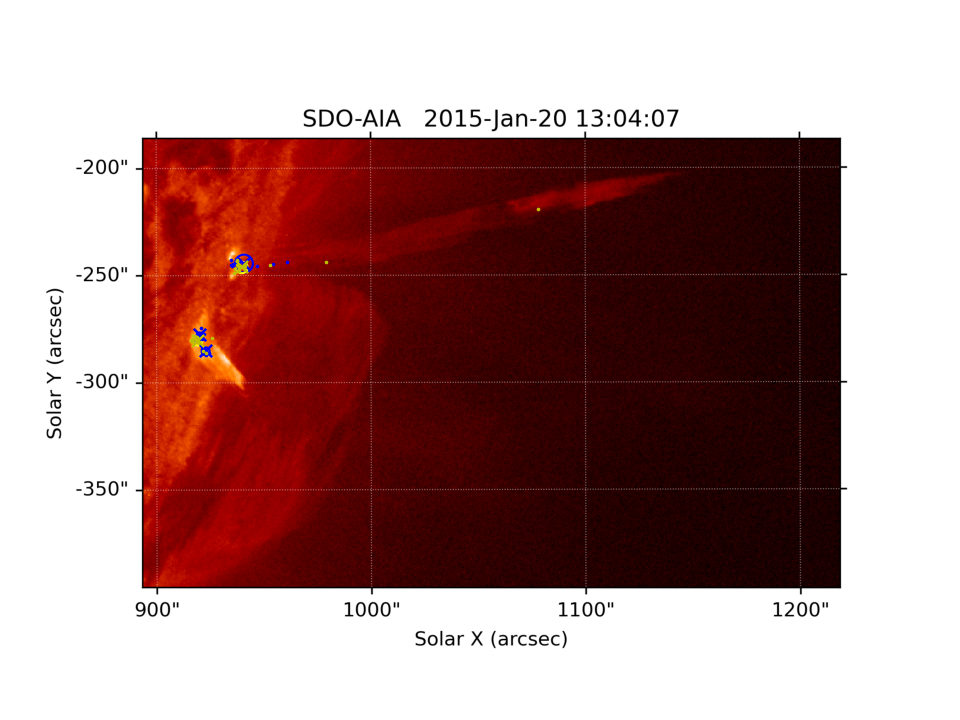}
    \caption{Point clustering for a subject. The blue points are the volunteer annotation of the base of the jet when it starts and the yellow points are the base of the jet when it ends. The blue and the yellow crosses are the corresponding aggregated cluster centers, and the circles denote the 1$\sigma$ error in the location of the cluster center. }
    \label{fig:point_clustering}
\end{figure}

\subsubsection{Clustering and aggregation of bounding boxes} \label{sec:box_clustering}

Volunteer annotations of boxes are done using the \texttt{rotateRectangle} tool on Zooniverse, which allows volunteers to draw a rectangle and rotate it to line up with the jet. As such, each bounding box is described using a corner location (x, y) on the image, a width, a height and an angle with respect to the vertical.  Similarly to the base locations, we use a density-based clustering algorithm to cluster and aggregate the boxes, but use DBSCAN \citep{Ester1996} instead of HDBSCAN, since the hierarchical nature of the HDBSCAN tended to favor one large uncertain cluster per subject rather than smaller densely packed clusters. 
Rather than cluster directly on the box parameters (since this is generally noisy and unstable, and also leads to issues with angular periodicity), we define the Jaccard metric to identify `distances' between boxes. This is based on the calculation of the Intersection-over-Union (IoU), which is the fraction of the total intersection area between any two boxes A and B and their total union area, and is given by,

\begin{equation}
    \textrm{IoU} = \dfrac{|A \cap B|}{|A \cup B|},
\end{equation}
where $|X|$ is the area of the collection of points in X.
In this way, two boxes that have perfect overlap have an IoU of 1, while two boxes that are completely separate have an IoU of 0. The Jaccard metric inverts this, so that a large overlap has a `distance' close to 0, while a smaller overlap has a `distance' closer to 1, given by,

\begin{equation}
    \textrm{Jaccard} = 1 - \textrm{IoU}.
\end{equation}

Unlike the HDBSCAN algorithm, the DBSCAN algorithm uses a flat cut to define clusters, given by the parameter $\epsilon$, which defines the maximum distance between boxes that are within the same cluster.
We tested with $\epsilon$ values of $0.3-0.7$ and found that a value of $0.6$ (i.e. a minimum of 40\% of the total area overlaps between any two volunteer annotations to be treated as the same jet) gave us good performance.

Finally, to get an aggregated (i.e., `average') box per cluster, we use the SHGO algorithm \citep{Endres2018} to optimize the average box parameter that minimizes the sum Jaccard distance between the average box and all volunteer annotations of that cluster. We calculate the 1$\sigma$ confidence in the average box parameters, $\sigma_{IoU}$ from the fit residual,
\begin{equation}
    \label{equation:sigma_iou}
    \sigma_{IoU} = \sqrt{\dfrac{\Delta}{N - 1}},
\end{equation}
where $\Delta$ is the residual and $N = 5$ is the number of parameters defining the shape (for boxes, these are the corner pixel $x$ and $y$ positions, width, height, and angle). An example of box clustering is shown in Figure~\ref{fig:box_clustering}. We detail the performance of our clustering algorithm in Appendix~\ref{appendix:box_clustering}.

\begin{figure}
    \centering
    \includegraphics[width=\columnwidth]{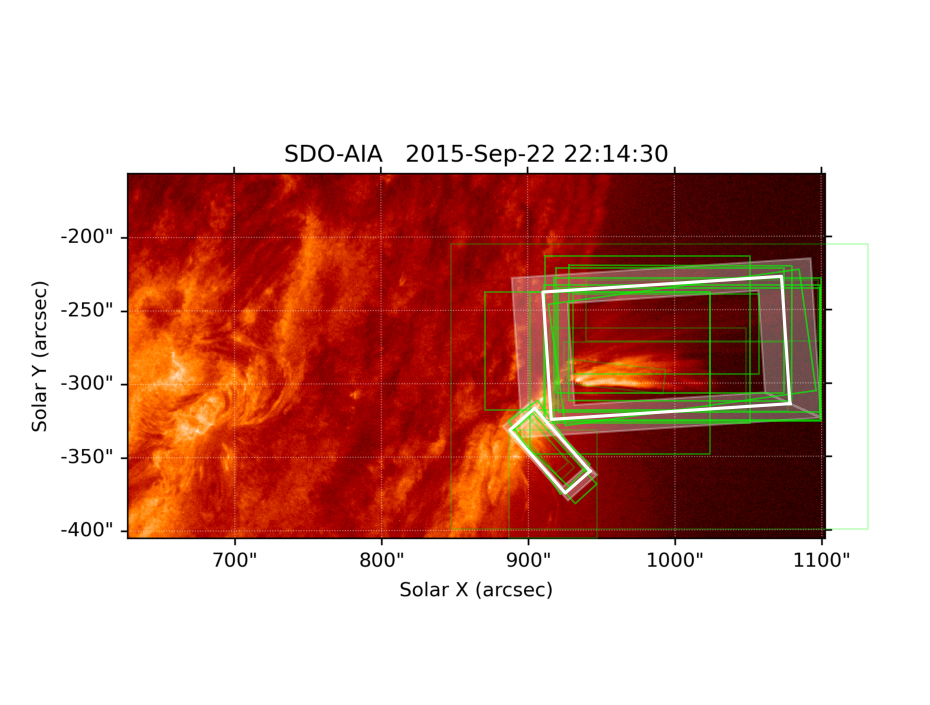}
    \caption{An example of the clustering of volunteer annotations into an aggregate result. The green boxes correspond to the volunteer annotations and the white boxes show the corresponding `average'. The white shaded region denotes the 1$\sigma$ uncertainty in each average box.}
    \label{fig:box_clustering}
\end{figure}

\subsection{Extracting multiple jets per subjects}

\begin{figure*}
    \centering
    \includegraphics[width=\textwidth]{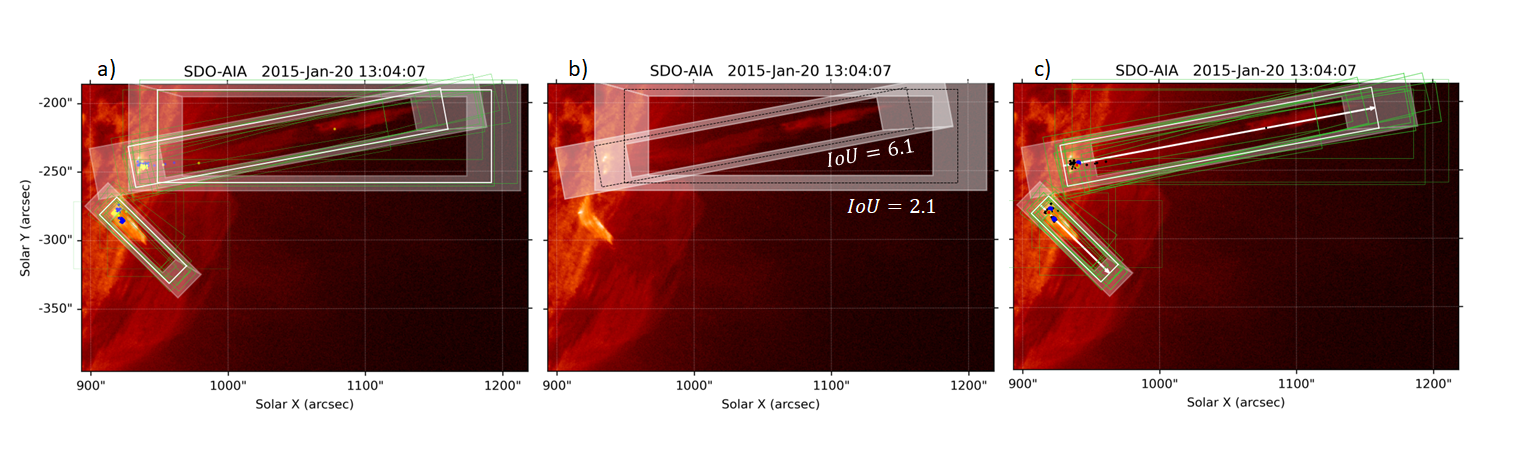}
    \caption{Process of cleaning up overlapping boxes and points. Panel (a) shows the image with two unique jets in the image, but three boxes. In panel (b), first, the overlapping boxes are identified and for each box cluster, we calculate the total IoU between the average, clustered box and the corresponding classifications that make up the cluster. In Panel (c), the box with the highest total classification IoU is chosen, and the rest of the averaged boxes are discarded.}
    \label{fig:unique_jets}
\end{figure*}

There are also cases where volunteer annotations are aggregated into overlapping boxes for a given jet, see e.g., Figure~\ref{fig:unique_jets} (a). To overcome this issue, we go through each subject and identify boxes that have significant overlap. Within this subset of boxes, we retain the box with the highest IoU with respect to the corresponding volunteer classifications and discard the remaining boxes. As seen in Figure~\ref{fig:box_clustering}, there is the possibility of having multiple jets within a given subject. Given that the boxes do not share a significant overlap, these boxes are not discarded and these result in two unique jet boxes. Similarly, there can be several start and end base points associated to a box. We calculate the mean distance between the start and end bases with the chosen `best' box as defined above, and find the respective base with smallest Cartesian distance from all edges of the box. As such, this allows us to build a set of unique boxes, with a corresponding start and end base which defines unique jets in the subject. This process is shown in Figure~\ref{fig:unique_jets}.

\subsection{From subjects to jets}\label{sec:subject_to_jets}

Each jet object within one subject is described by one box and two base locations, one at the first frame that it appeared in and one at the last frame that the jet was visible in. The box is only described by the coordinates of the four corners and does not contain any information about the orientation and direction of the jet. In order to differentiate the height from the width, we calculate the distance between each corner of the box and the base of the jet reported by the volunteers. 
The width of the box is identified as the edge of the box between the two corners that are closest to the base. The height is then the perpendicular edge of the box.

In order to track the jet over multiple subjects, we perform spatial and temporal clustering (steps 3 and 4 in Figure \ref{fig:data}). This clustering is done in several steps:
\begin{enumerate}
    \item First, we determine all the subjects that correspond to a single HEK event. We collect all the unique jet boxes and start and end bases for these subjects.
    \item We do a spatial clustering on both the boxes and bases. We do a simple cut on the IoU greater than 0.1 for the boxes, and use the Euclidian distance for the bases to determine which subjects are part of the same cluster. The dimensionless distance between two jets $A$ and $B$ is defined as,
    \begin{equation}
        d = 2 \times {\rm Jaccard(A, B)} + \frac{d_{b, AB}}{d_{b, 90}},
    \end{equation}
    where $d_b$ is the sum of the distance between the start ($s$) and end ($e$) bases, $d_{b, AB} = |s_{A} - s_B| + |e_A - e_B|$, and $d_{b,90}$ is the $90$th percentile of $d_b$ for all jets in the HEK event, used for normalizing the pixel distances. The factor of $2$ in the Jaccard metric is to ensure that the point and box metric are scaled similarly (i.e., so that one metric is not weighted more than the other). We tested different percentile values to normalize the base metric, and found that the $90$th percentile produced robust results. When $d < \epsilon_s$, we consider the jets to be spatially clustered. We tested different values of $\epsilon_s$ and found that a value of $3$ produced the best clusters across all the subjects.
    \item For each spatial cluster, we further filter the boxes and points based on their time stamp. We put a requirement that the annotations belonging to the same jet from multiple subjects must have time adjacency, i.e., we constrain that a jet in a given subject (which is part of a spatial cluster defined above) must be at most 2 subjects from a jet within the same spatial cluster. Physically, this means that two annotations must be at most 10m 24s apart to be considered part of the same jet.
\end{enumerate}
The result of this spatial clustering and time filtering are jets spanning over different subjects of the original HEK event reported, as illustrated in Figure \ref{fig:data}. 
Additional filtering of the result is applied. First, we require the short jets that are reported in only one subject to have at least 50$\%$ agreement on the first question of the second workflow, which is a confirmation that this subject contains a jet. Secondly, we remove the jets with a high uncertainty on their associated box by keeping only jets with an $\sigma_{IoU}<0.8$ in our final sample.

\section{Results}
\label{sec:results}

\subsection{Engagement in the project}
\label{sec:results-enagagement}

After two sessions of beta-test, Solar Jet Hunter was publicly launched on December 7, 2021. The data set used contained the coronal jet events reported in the HEK in years 2011, 2012 and 2013. On December 7, 5205 classifications were performed, and on December 8, 6040 classifications were performed. With such a high number of classifications, the first workflow "Jet or Not" was completed in less than three days. Jets were found in 31\% of the data in that workflow. These 31\% of the subjects were sent to the second workflow, "Box the Jets". This second workflow requires a deeper involvement from the volunteers, and the retirement limit is higher, so the analysis of the data in this workflow took more time but was still completed in less than two months. 

After this first run of the project, we relaunched it twice. The second launch was on April 22, 2022 with data from 2014 and 2015. This data set contained 2897 subjects, and the classification was completed at the beginning of August. The third launch happened on October 11, 2022 with data from 2016, which consist of 2831 subjects. The classification was completed in February 2023. At the end of the year 2022, more than 3,000 volunteers were registered and more than 120,000 classifications had been performed in the project.

The aggregation of the results was performed as described in section \ref{sec:aggregation}. After the clustering and filtering procedures described in section \ref{sec:subject_to_jets}, jets were found and boxed in only 21\% of the original data set. This suggests that 79\% of the data reported as belonging to a "coronal jet" event in HEK does not contain a jet. Since we found 883 jets in data originating from 364 "coronal jet" events in HEK, it also suggests that many HEK events contain multiple jets. 

As mentioned in section \ref{sec:project-setup}, one important aspect of the Zooniverse is the Talk board where discussions can happen between volunteers, and with the research team. In particular, specific subjects can be discussed. At the end of 2022, more than 1,200 individual subjects have been discussed by the volunteers. A small fraction of the volunteers (around 160 volunteers, $\sim$ 5\% of the registered volunteers) participate in these discussions. The number of classifications seems to be related to the reactivity of the research team to answer these messages, provide guidance or additional information, both about the classification process, but also about the overall goals of the project and the global questions in solar physics.

\begin{figure}
    \centering
    \includegraphics[width=\columnwidth]{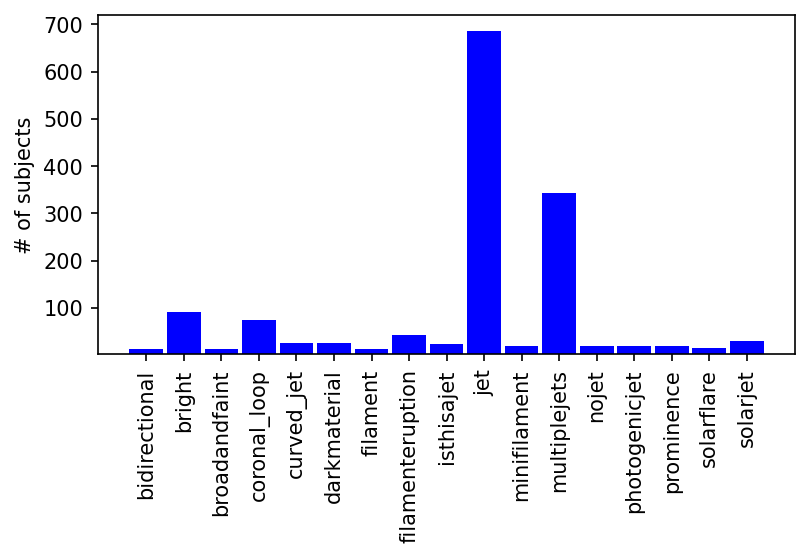}
    \caption{Distribution of tags used in the Talk board by volunteers. }
    \label{fig:talk_tags}
\end{figure}

In the workflow "Box the Jets", the volunteers are asked to post a comment in the Talk Board for subjects in which they found more than two jets, and to tag the discussion with \emph{\#multiplejets}. This tag has been used for 341 subjects so far. Additionally, volunteers spontaneously created and used other tags, such as \emph{\#jets}, \emph{\#isthisajet}, \emph{\#minifilament}, \emph{\#bright}, \emph{\#prominence}, \emph{\#curved-jet}, \emph{\#bidirectional}, to cite some of the popular ones. Figure~\ref{fig:talk_tags} shows the distribution of some of the most popular tags used in the Talk board. Given that volunteers can provide different spellings for the same tag (e.g., \emph{\#multijet} vs \emph{\#multiplejet} vs \emph{\#multiplejets}), we combine these multiple tags manually. These tags are added values to the classification and enable the researchers to browse for specific data in the Zooniverse. For instance, after a discussion with a researcher of the team in the Talk Board about mini-filaments, the tag \emph{\#minifilament} was used by volunteers to report on some jets for which they thought they detected a mini-filament eruption. The tag \emph{\#spinning-jet} was also spontaneously created by volunteers to tag jets in which they saw what researchers call "untwisting" motion. This tag, which arose without prompting, evokes the physical distinction between "straight" and "helical" jets arising in the magnetohydrodynamic simulations of \citet{pariat2015}. The tag \emph{\#photogenicjet} is also useful to pick subjects particularly pleasing visually, for instance to promote the project. The spontaneous creation and use of these tags are one illustration of the way volunteers often go beyond their task in citizen science project, and could drive serendipitous discoveries. 

\subsection{Catalogue of coronal jets}

The full catalogue of jets resulting from Solar Jet Hunter, together with information gathered during the aggregation, clustering and filtering, is available as a JSON database as well as a reduced version of this in csv: both are available at \url{https://conservancy.umn.edu/handle/11299/257209}. Both contain the following jet properties for each jet: the start and end times, duration, base location, in heliocentric coordinates and in Stonyhurst heliographic coordinates, maximum height, average width, ID of the corresponding HEK event. When available, the estimated jet velocity is provided. The jet sigma ($\sigma_{IoU}$) and flags attributed to the jets are also provided. In the following we describe how these parameters have been calculated.

The duration of a jet is calculated as the difference between the time stamp of the last image of the last subject and the time stamp of the first image of the first subject in which the jet was detected. When the jet is longer than one subject we calculate the average and standard deviation of the start base over each subject involved in this jet. We additionally calculate the average and standard deviation of the jet width. For the height we take the subject for which the maximum value is achieved and also calculate the lower bound and upper bound of this height, these values being obtained by using the $\sigma_{IoU}$ uncertainty of the box in that given subject. By taking the maximum height divided by the time at which the maximum height is reached minus the start time, we get a rough estimate of the average velocity of the jet projected onto the plane of the sky in km/s. 
Jet entries in the catalogue are also associated with quality flags. A binary flagging system is used, with flag 000 in the case of no quality flag. Flag 100 is given to a jet if it lasts for less than 6 minutes; flag 010 is given if we have a longitude above the 90$^{\circ}$for the base point, thus base points unexpectedly elevated above the solar surface; and flag 001 is given to a jet if the averaged velocity could not be calculated.

\subsection{Statistics on the catalogue entries}
\label{sec:stat}

In this section, we present the statistics as they are present in our catalogue, based on 883 jets. In table\,\ref{tab:statsjets} the mean, standard deviation, median, minimal and maximal values of the jets heights, widths, durations and velocities are presented. The height shows a higher average and median than the width and a higher standard deviation, fitting with the view of jet ejections being collimated. The boxes of the jet clusters start small and later stretch to a larger height while the width remains the same. 

\begin{table}
    \centering
    \begin{tabular}{c c c c c c }
    \hline
    & Mean & Std Dev & Median & Min & Max \\ \hline
H$^{304\,\AA}$ &  121 & 77 & 100 & 15 & 434 \\
W$^{304\,\AA}$ & 52 & 44 & 39 & 4 & 383 \\
D$^{304\,\AA}$ & 18 & 18 & 13 & 5 & 182 \\
V$^{304\,\AA}$ & 160 & 138 & 131 & 15 & 1702 \\
\hline
    \end{tabular}
    \caption{Statistic of heights, widths, durations and velocities of jet clusters found from AIA 304\,$\AA$ images: mean, standard deviation, median, minimal and maximal values.}
    \label{tab:statsjets}
\end{table}

\begin{figure}
    \centering
    \includegraphics[width=0.95\linewidth, trim=0.0 8 0.0 10.5, clip]{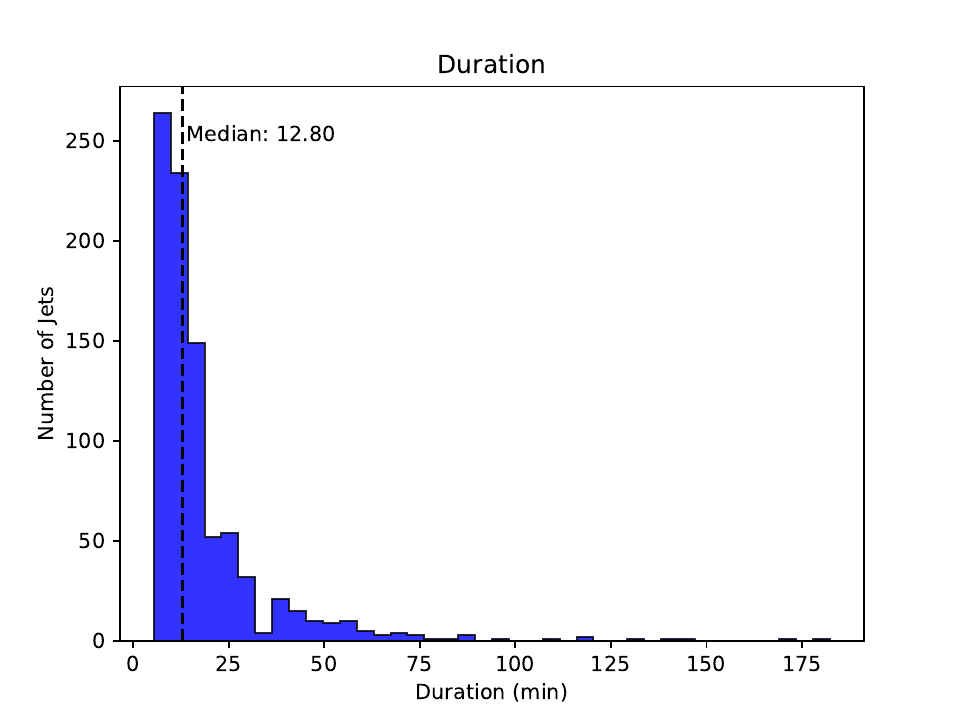}
    \includegraphics[width=0.95\linewidth, trim=0.0 8 0.0 10.5, clip]{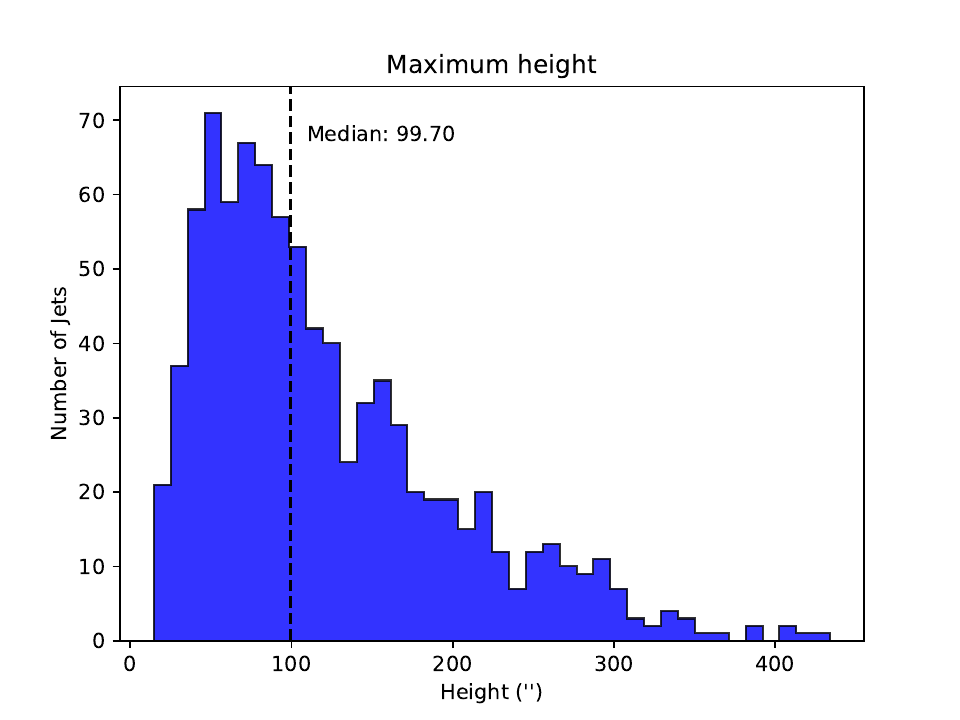}
    \includegraphics[width=0.95\linewidth, trim=0.0 8 0.0 10.5, clip]{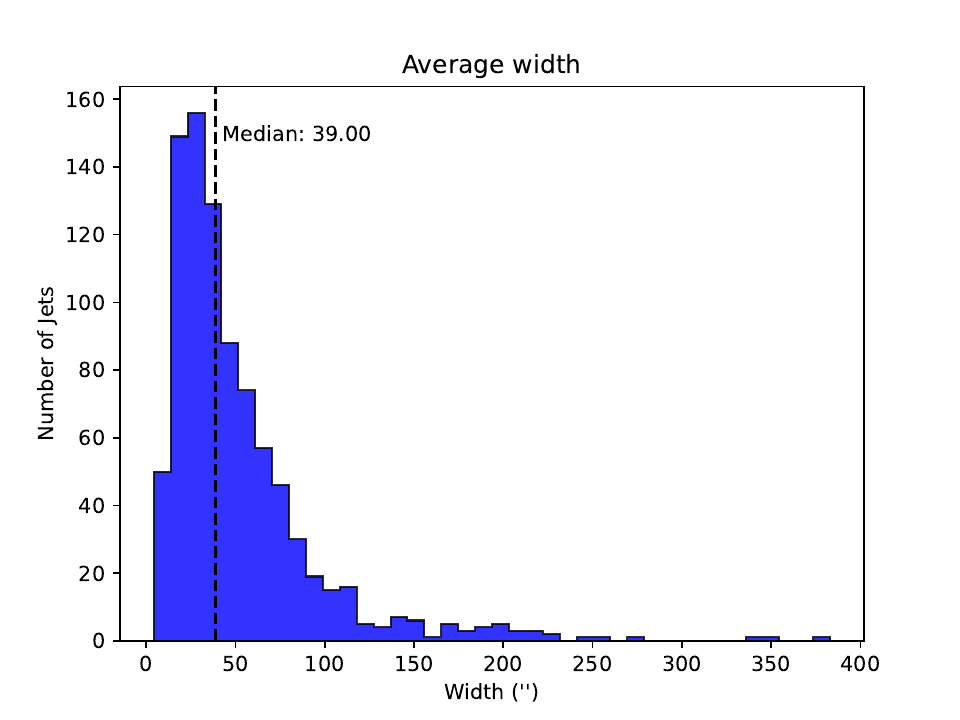}
    \includegraphics[width=0.95\linewidth, trim=0.0 8 0.0 10.5, clip]{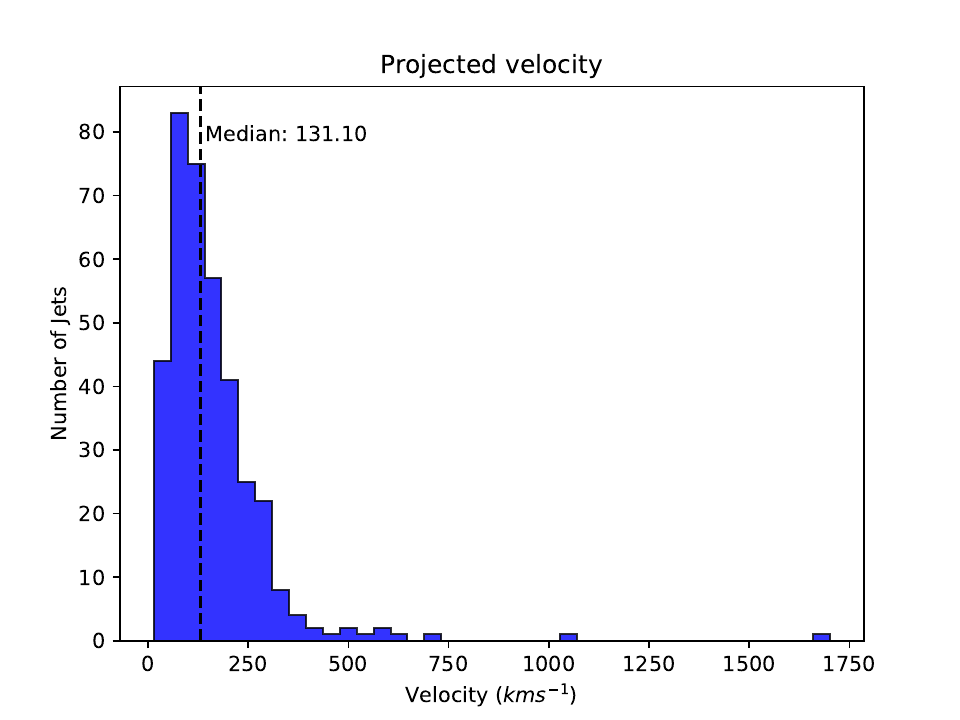}
    \caption{Distributions of jet properties for the 883 jets in the catalogue, from top to bottom: duration of jets in minutes, maximum height measured in arcsec, average width in arcsec, and estimated projected velocity in km/s (only for a subset of jets, see text). }
    \label{fig:stat_cat}
\end{figure}

The distribution of jets' duration, in the top panel of Figure\,\ref{fig:stat_cat}, shows that a large number of short jets are present in the catalogue. Because of the way the jet duration is calculated, the minimal duration we measured is the duration of one subject (6 minutes). A large number of jets with a duration of 6 minutes is present in the sample, as seen in Figure\,\ref{fig:stat_cat}. A large number of short jets of 6 minutes is unexpected based on previous research, however, jets of this duration have been observed before. In \cite{musset2020} a minimal value of 4.2 minutes is given for their 33 jet sample. 

The next two panels in Figure\,\ref{fig:stat_cat} show the distributions of the projected height and width of the jets, respectively. The distribution of the width shows a drop-off as close as 150 arcsec. The height on the other hand seems to decrease gradually until 350 arcsec, with some higher values outside of the main distribution in both cases. The median value of the height is 100 arcsec, higher than the median value of 39 arcsec found in the width.

The velocity estimate histogram is only computed for subjects that do not have their maximum height in the first subject and thus is composed of fewer jet clusters than the other statistics (371 jets). The result is shown in the bottom panel of Figure\,\ref{fig:stat_cat}. The median value is around 131 km/s with most values being between 15 and 350 km/s. In our calculation we take the maximum height divided by the time needed to reach that height, meaning we are averaging over time, which could lead to lower values. The velocity estimate of the jet clusters in the catalogue should only serve as a crude indication of the jet velocity.

\begin{figure}
\includegraphics[width=0.98\linewidth]{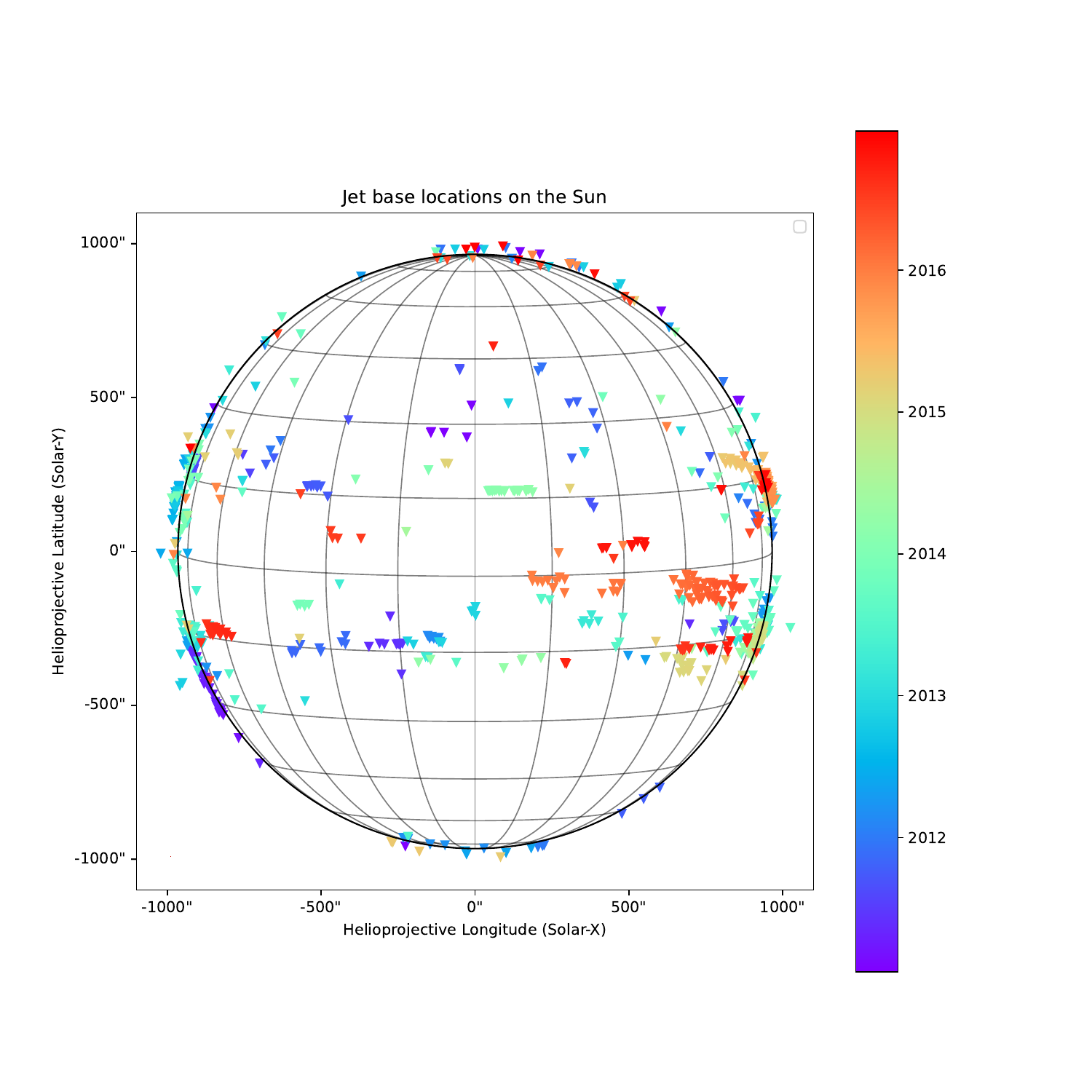}
\caption{The base point locations of the 883 jet clusters between the years of 2011-2016. The vertical axis and horizontal axis show the Helioprojective X and Y coordinates, respectively.}
\label{fig:solarmap}
\end{figure}

Figure\,\ref{fig:solarmap} shows the visual representation of the detected jet clusters base point locations on the Sun map during detection. Jets that are detected on disk  are distributed around  $\pm$ 15 degrees in latitude, or at the poles. This distribution matches the distribution of active regions were numerous jets are found, and confirm that many jets are also found in polar regions, most probably associated with coronal holes boundaries. Jets are also concentrated at the limb of the solar disk, confirming that it is easier to detect jets by eye at the limb rather than on the disk, in the 304 \AA \ observations. Finally, it is clear on this figure that jets tend to cluster in time and space, confirming that regions producing jets will produce often produce a whole series of jets.

\section{Discussion}

\subsection{Comparison with previous catalogues}

\subsubsection{Comparison with the Heliophysics Events Knowledgebase (HEK) }

The Heliophysics Events Knowledgebase is the most extended database of coronal jets to date. However, as mentioned in section \ref{sec:project-setup}, events reported in this database do not provide precise information on individual coronal jets. This is confirmed by the results presented in
section \ref{sec:results-enagagement}. The events reported in HEK represented the input for the Solar Jet Hunter. For the coronal jets reported in HEK in years 2011 to 2016, jets were found by the volunteers in only 21\% of the data. We also found a total of 883 jets within the 364 events reported in HEK. Our catalogue reports individual jets, and the time intervals reported in this catalogue are the times at which the jet was actually present in the data, which is a notable improvement on the HEK database.

\subsubsection{Comparison with previous statistics on coronal jets}

A few statistical studies of jets exist and reported on jet properties. For instance, \cite{panesar_2016a} examined 13 homologous jets happening at the edge of an active region with the SDO/AIA instrument, \cite{mulay_etal_2016} examined a sample of 20 active region jets using different EUV passbands of SDO/AIA, and \cite{Joshi_etal_2017} examined the velocities of 18 jets recurring on one active region, using the 171 \AA \ channel of SDO/AIA. \cite{musset2020} reported statistics of jets parameters using a sample of 33 jets examined in the 304 \AA \ EUV channel of SDO/AIA. Using their recent catalogue of coronal jets observed by SDO/AIA, \cite{Anfinogentov_etal_2021} and \cite{Stupishin_etal_2021} performed a preliminary statistical analysis of 80 jets to report on their timing and size.
\cite{odermatt_etal_2022} performed an analysis of 180 jets found in 5 active regions, and measured the length and velocity of these jets taking into account the curve of the jets in the plane of sky, using the 193 \AA \ EUV channel of SDO/AIA. We note that most of these studies focus of active region jets, and that \cite{mulay_etal_2016} and \cite{musset2020} focus specifically on flare-associated jets.

\begin{figure}
\includegraphics[width=0.98\linewidth]{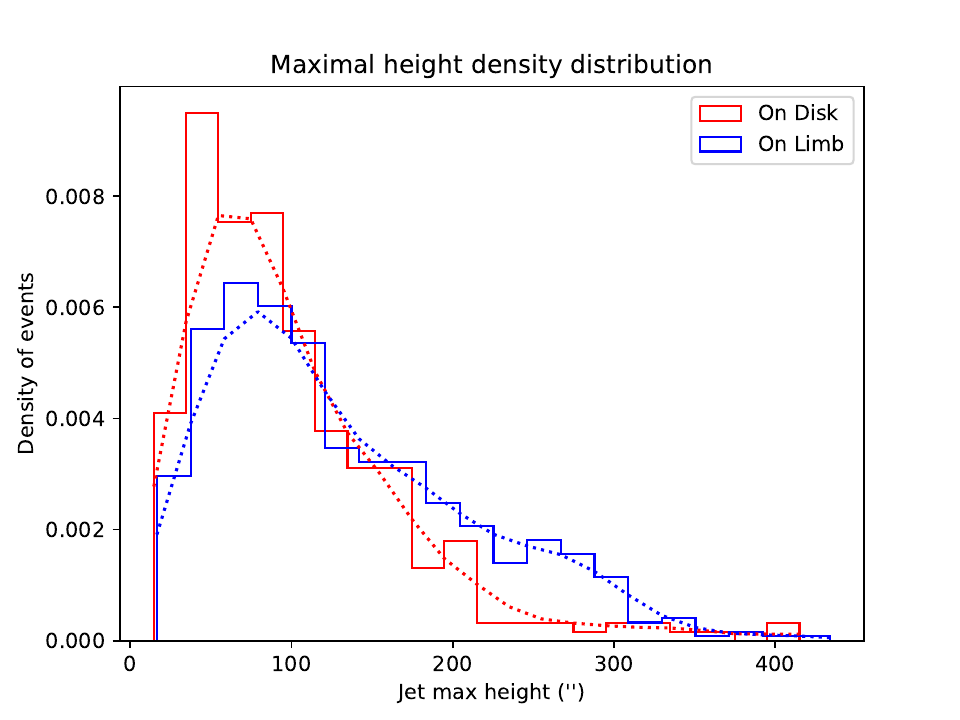}
\caption{Density distribution of jet maximum height for on-disk (radial distance from disk center lower than 900 arcsec) and on-limb (radial distance from disk center greater than 900 arcsec) jets.}
\label{fig:disk-limb}
\end{figure}

The elongation of the jets reported by \cite{Stupishin_etal_2021} ranges between 13 and 144 arcsec, with a mean of 89 arcsecs. Most of the jets reported by \cite{odermatt_etal_2022} have lengths smaller than 70 arcsec, with some longer jets reaching 130 arcsec in length. Since the study by \cite{odermatt_etal_2022} takes into account the curvature of the jets, we expect the lengths reported to be in average higher than the maximum height of the jets reported in other past studies, and from Solar Jet Hunter, where lengths are measured as the height of the longest box. However, the maximum height of the jets reported in our study as a median of 100 arcsecs, and jets as long as 400 arcsec have been reported. Since \cite{odermatt_etal_2022} limited their sample to jets seen on-disk, we explored the possibility of this difference being due to strong projection effect for on-disk jets. We divided our sample in two categories: on-disk and on-limb jets, with the criteria of their radial distance to the solar disk center being below or above 900 arcsec respectively. The distributions of maximum heights for both sub-samples is shown in Figure \ref{fig:disk-limb}. The distributions in heights are similar, with a slight difference that would tend to confirm that the height of on-disk jets tend to be smaller than for jets seen at the limb.

\cite{mulay_etal_2016} reported jet lifetimes (duration) between 8 and 39 minutes, with a median of 20 minutes, and \cite{panesar_2016a} reported lifetimes 10 and 50 minutes. \cite{musset2020} reported a median jet duration of 18.8 minutes in their sample. These studies are reporting on active region jets. \cite{Stupishin_etal_2021} used a catalog of jets which should contain jets in and outside of active regions, and reported the the jet duration of their sample of 80 jets lied between 1 and 17 minutes, suggesting that including jets from outside of active regions results in the inclusion of jets with shorter lifetimes.
The duration of jets reported by Solar Jet Hunter is thus consistent with these reports and other studies, and in particular contains jets with short time scales which could be jets outside of active regions. On the other hand, the great number of short jets found by the citizen scientists might be the result of a methodological biais, further discussed in the next section.

The velocity reported in our catalogue is only a crude estimate, because no time-distance analysis was performed, so no significant quantitative conclusion can be derived from the distribution we report. However, the numbers derived can still be compared to past statistical studies that computed jet velocities.
The average jet velocity of 163 km/s in our catalogue is consistent with the average velocities found in previous statistical studies of jets, in which the velocity was calculated with a height-time analysis: 156 km/s \citep{Joshi_etal_2017}, 198 km/s \citep{panesar_2016a}, 207 km/s \citep{musset2020}, 271 km/s \citep{mulay_etal_2016}. We note that these last two studies with higher mean velocities focused on jet associated with flares, maybe suggesting that flare-associated jets are faster than jets not associated with flares. \cite{panesar_2016a} already noted that jets associated with CMEs were significantly faster (in average, 300 km/s) than jets that were not associated with CMEs (105 km/s in average). On the other hand, \cite{musset2020} found no correlation between jet velocities and flare intensity in their sample.
\cite{odermatt_etal_2022} calculated the jet velocities differently compared to the studies cited so far, and found that most of their jets have a velocity below 200 km/s with a median at 114 km/s. The velocities reported in the Solar Jet Hunter catalogue also compare with these numbers with a median value at 131 km/s.

\subsection{Methodology bias}

We identified several bias in our methodology.
First, we so far only used time intervals and regions of the solar disk that already have been reported in the HEK database as jets. Therefore, any jet missing from the HEK database will also be missing from our sample. This will be addressed in the future by analysing the whole AIA dataset, which will be done either with machine learning algorithms or during an extension of the citizen science project. 
Second, we restricted our analysis to 304 \AA . This can also be addressed in an extension of the citizen science project, using the catalogue created with 304 \AA , and showing the data in other wavelengths to volunteers for them to report in which wavelengths the jets is visible. 

Solar jet hunter is a project in which we ask the volunteers to find jets. Therefore, volunteers may be tempted to report jets in subjects where a researcher would not have taken the effort, for instance if the moving feature is very short or very small. While this is positive in the sense that many small and short events will be reported, this  may be pushing the definition of jets to small scales in which events were often reported as "jetlets" \citep{raouafi14,panesar18b}.  
This points to a blurry transition between what solar physicists call "jets" and tinier events. This bias is not necessarily strong as our data are analyzed by many different volunteers, reducing the noise or the bias linked to the "human factor" (the fact that one volunteer may tend to report tiny events while another will only report large and obvious events). On the other hand, it could explained the large number of jets with durations below 6 minutes that were reported. This bias will be tested during the next runs of the project, with video subjects that will cover longer time intervals than the 6-minute movie strip that was used so far (see next section).

Our choice of a 24 sec cadence is also a limiting factor on the duration of the jet events that will be reported by this project.

\begin{figure*}
\includegraphics[width=0.99\linewidth]{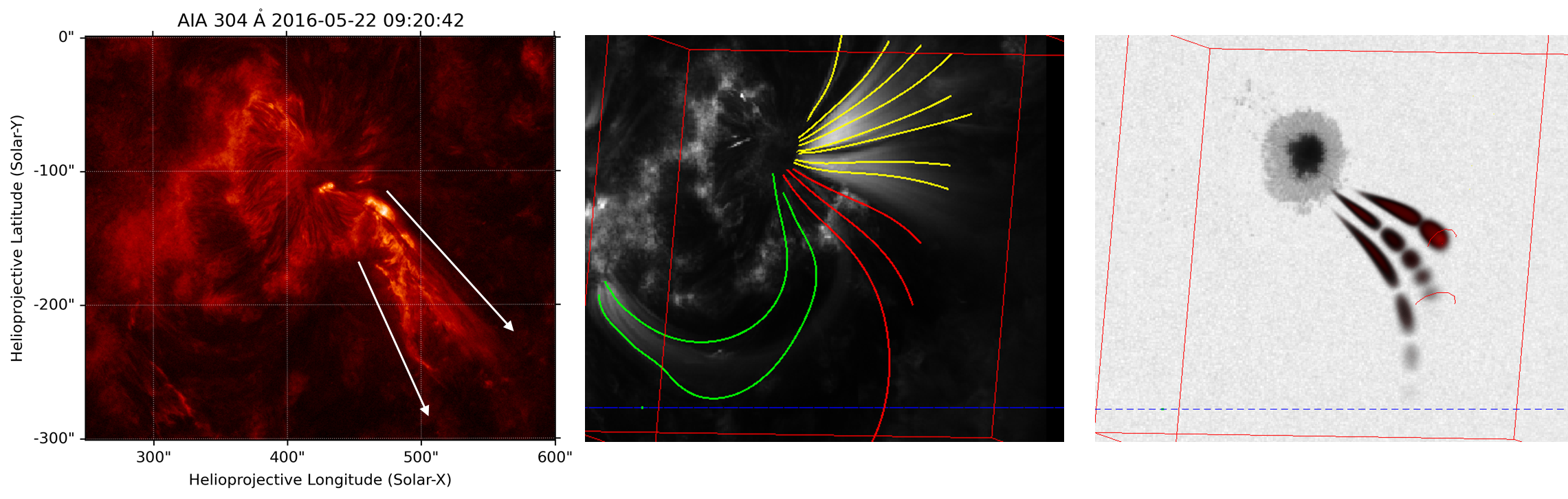}
\caption{Example of magnetic connectivity analysis for a jet in our catalogue using NLFFF extrapolation. Left: jet detected in 304 \AA, highlighted by the two white arrows. Middle: magnetic connectivity in the NLFFF data cube: open magnetic field lines are shown in yellow and closed field lines are shown in green on top of the the 171\,\AA\ image. These lines are consistent with the the 171\,\AA\ bright features including remote bright loops on the left outlined by the green lines, which confirms overall validity the NLFFF model. The open field lines at the cold (304\,\AA\ passband) jet location are shown in red. They do not have any obvious hot counterpart in the 171\,\AA\ passband. Right:   idealized blobs of ejected cool thermal plasma moving away from the sunspot along the red field lines are show as dark red volume.
}
\label{fig:extrapolation}
\end{figure*}

\subsection{Future steps}

\subsubsection{Improvements and extensions of the citizen science project}

There are numerous ways to upgrade and extend this project in order to continue and improve the detection of solar jets in EUV data. We describe in the following a few avenues that we are considering for Solar Jet Hunter: the use of video tools, the introduction of different levels of difficulty for subjects, the extension of the input beyond the HEK catalogue, and the use of machine learning in synergy with the citizen science project.

The citizen science project Solar Jet Hunter will be upgraded and extended in several ways. A first upgrade will be achieved in 2023 with the introduction of video subjects instead of movie strips. Annotating videos is a new capability of the Zooniverse framework. Replacing our current subjects with video subjects allows us to propose videos with around one hundred frames instead of fifteen. With the cadence of 24 seconds chosen for the project, this means that volunteers will see 20 minutes of solar data in one subject, instead of six minutes: it might facilitate the task of finding jets in the data. On the aggregation point of view, this will leave enough statistics to cluster the volunteers annotation in time as well as in space, ultimately resulting in finer time information in the jet catalogue.

Another way to upgrade the project would be to assign the "difficult" subjects only to experimented volunteers in order to improve the accuracy of the classification.
As mentioned in section \ref{sec:aggregaion-volunteers}, we can use the agreement defined by equation \ref{equation:agreement} to evaluate the difficulty of a subject.
This can be used to assign subjects to different levels of difficulties. In parallel, the skill of a volunteer in detecting jets can be measured by comparing their answer to the result of the aggregation. With this two measurements, we could assign the most difficult subjects to the most skilled volunteers: in practice, volunteers reaching a certain skill level would unlock a workflow previously hidden, containing the most difficult subjects. This functionality will be tested for Solar Jet Hunter in the future.

As already mentioned, the current project input are the "coronal jet" entries from the HEK database. However, prominent jets have been found to be missing this database, so it is likely that numerous jets are escaping our detection. On the other hand, as described in section \ref{sec:stat}, jets tend to happen in clusters, or in other words, regions that are producing one jets are likely to produce other jets. One extension of the project could be to produce new data sets, targeting the regions in which we already found jets, but at different times. One could also use events flagged by an automated algorithm detecting flows in the data as inputs, to ask the volunteers to confirm the detection of a jet. Alternatively, one could decide to feed all of the data from recorded by AIA in 304\,\AA \ into the project, but in doing so, the probability of finding a jet in a given subject will be largely decreased. Finally, this project could be extended with data from other AIA channels and from other instruments than AIA.

The citizen science project can also be extended with other tasks, and maybe more involved tasks. For instance, the jets for which no clear agreement has been found in the first run of the Solar Jet Hunter could be shown again to the volunteers with the result of the aggregation, for validation. Subjects that have been tagged as containing multiple jets could be part of a new workflow in which all the jets are boxed, not only the two most prominent.

Futhermore, we are working on implementing a machine-learning based framework to detect coronal jets. We are working on using the current set of classifications to train a region-proposal based model \citep[e.g., Mask R-CNN,][]{MaskRCNN} to draw a bounding box around the jet in the images. Recently, transformer-based models \citep{Vaswani2017} have shown great promise in image recognition tasks, particularly in the avenue of object detection and instance segmentation \citep{Dosovitskiy2020}. We are exploring these avenues for training a region proposal model in an effort to accelerate the search for coronal jets. 

\subsubsection{Exploitation of the jet catalogue}

The jet catalogue presented in this paper is available to the scientific community as a tool to select coronal jet events for statistical and event-based studies. Careful analysis of jets in this catalogue is beyond the scope of the paper. For instance, the careful calculation of the jet velocity is left to further investigation. Jets reported in this catalogue correspond to a loose definition of coronal jets: collimated ejections of plasma seen in the data. A number of jets follow a curved trajectory as if they were propagating along a close loop in the corona rather than being ejected on an open magnetic field line. 

We believe that the same physical processes may be driving the jets whether this ejection is following closed or open magnetic field line. However, to analyze how coronal jets impact the heliosphere, one might need carefully analyze the magnetic field configuration to only include jets on open field lines in one's study. An example of such analysis is shown in Figure \ref{fig:extrapolation}, where the jet, outlined by two white arrows, is seen in the right bottom corner of the left panel. To investigate the magnetic topology and connectivity, we created  a non-linear force-free field (NLFFF) extrapolation of the magnetic field measured at the photospheric level by the Helioseismic and Magnetic Imager (HMI) on SDO using the automated model production pipeline available within the GX Simulator distribution \citep{2023ApJS..267....6N}. The middle panel of Figure \ref{fig:extrapolation} shows a set of open and closed magnetic field lines, computed from the NLFFF 3D data cube overlaid on the AIA image at the 171\,\AA\ channel. It is evident that a subset of the open and closed field lines shown in yellow and green, respectively,
matches well bright EUV loops highlighting $\sim1$\,MK coronal plasma. This confirms the overall validity of the NLFFF data cube. Another subset of the open field lines shown in red does not have the corresponding 171\,\AA\ bright loops, but these lines correspond to the cool (304\,\AA) jet in the left panel. The right panel shows a snapshot of an idealized plasma ejection (dark red volume) along the three open field lines shown in red in the middle panel.  This injection consist of several consecutive plasma blobs moving away from the sunspot; specifically---from an umbra-penumbra boundary, which implies that the jet launch can be associated with a change of the magneto-convection regime between the umbra and penumbra.


Further analysis of jets could also be performed by motivated volunteers of the citizen science project willing to go beyond the tasks proposed in "Solar Jet Hunter". 
Indeed, in previous citizen science initiative, the possibility for volunteers to get involve beyond the regular task of the project sometimes led to interesting scientific discoveries by the volunteers. Deeper engagement can be facilitated by providing tools to explore the metadata associated with the subjects, to explore the catalogue of jets produced by this project, and to combine observations of jets in AIA with other data sets. Such tools with the volunteers of Zooniverse as a target audience requires the use of stable, user-friendly, open source/free tools. We will develop tools to facilitate the access to the jet catalogue produced by the project, and promote existing tools that are available and user-friendly to explore solar data sets, such as Helioviewer\footnote{\tiny \url{https://helioviewer.org/}} and JHelioviewer\footnote{\tiny \url{https://www.jhelioviewer.org/}}, in order to favor deeper engagement of the solar jet hunters.

\section{Conclusion}

In this paper we report the successful use of citizen science to detect and analyze solar coronal jets in the AIA 304 \AA \ data. We demonstrated here that there is a strong interest from the public to participate in the project with continuous engagement over several months. Solar Jet Hunters were able to deal with challenging data, such as the solar EUV images that contain complex features, and to manipulate both time and spatial dimensions in their analysis. This project therefore illustrates the use of citizen science to perform both the exploration and the first-order analysis of a large and complex data set: citizen science could therefore be a valid, and the most appropriate methodology, for other data sets in solar physics. Moreover, it participates in building a relationship between academia and the general public, promoting science, the scientific method, and scientific careers, and providing a platform for discussions and collaborations between academic and amateur scientists. 

The result is a catalogue of solar jets with a precision on timing and locations greatly improved compared to previous reports of coronal jets, with reliable error estimations. This catalogue is publicly available to the community and will be regularly enhanced with new events as the Solar Jet Hunter will continue to run. We envision this catalogue to be used to explore the questions of how jets are generated, how they are connected to particle acceleration and energetic particle injection in the heliosphere, and how they impact the solar wind.


\begin{acknowledgements}
The Solar Jet Hunter research team addresses many thanks to the Solar Jet Hunter volunteers for their interest in the project, their participation in classifications and discussions, and their suggestions for improvement. They also thank the Zooniverse team for their support both in preparation and during the project. Solar Jet Hunter is supported by the NASA grant 80NSSC20K0718, NASA grant 80NSSC20M0057 and NSF grant AGS1752268 and the Solar Jet Hunter team thanks NASA for their support of citizen science, the discussions and tools they provided, as well as their advertisement of the project.
S. Musset is supported by the ESA Research Fellowship programme.

\end{acknowledgements}

\appendix
\section{Clustering performance} \label{appendix:clustering}
In the following sections, we describe the performance of our clustering algorithms and detail the methods used to quantify the uncertainty in the final cluster averages.

\subsection{Point clustering} \label{appendix:point_clustering}
\begin{figure}
    \centering
    \includegraphics[width=\columnwidth]{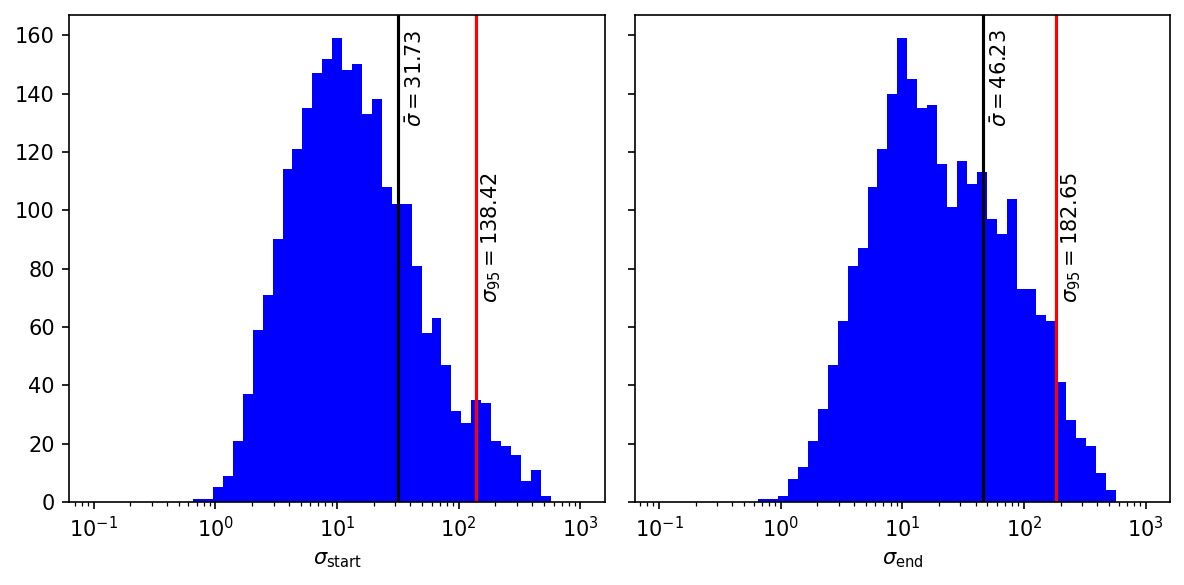}
    \caption{Distribution of $\sigma$ for the start and end base points for the jet. The start points are known with greater certainty with a mean of $31$ pixels, while the end points have a mean of $46$ pixels, due to volunteers annotating the end of the jet, rather than the base of the jet at the last frame that it is visible.}
    \label{fig:point_cluster_dist}
\end{figure}

We cluster the base of the jet at the start and end frames using the HDBSCAN algorithm with the Euclidian distance, and the aggregated point is the center of the cluster. As such, we define the uncertainty in the cluster average as the mean distance between the cluster average and individual points that define the cluster. Figure~\ref{fig:point_cluster_dist} shows the distribution of uncertainties in point clusters across all the jets found by volunteers, along with the corresponding mean and 95th percentile values. The mean uncertainty in the point cluster was about 33 pixels for the start of the jet, while about 47 pixels for the end of the jet. In general, this corresponds to about 7 arcsec and 11 in the uncertainty for the start and end points, respectively. The end points were more challenging to cluster, primarily due to the volunteers marking the end of the jet rather than the base of the jet at its ending frame.

\begin{figure}
    \centering
    \includegraphics[width=\columnwidth]{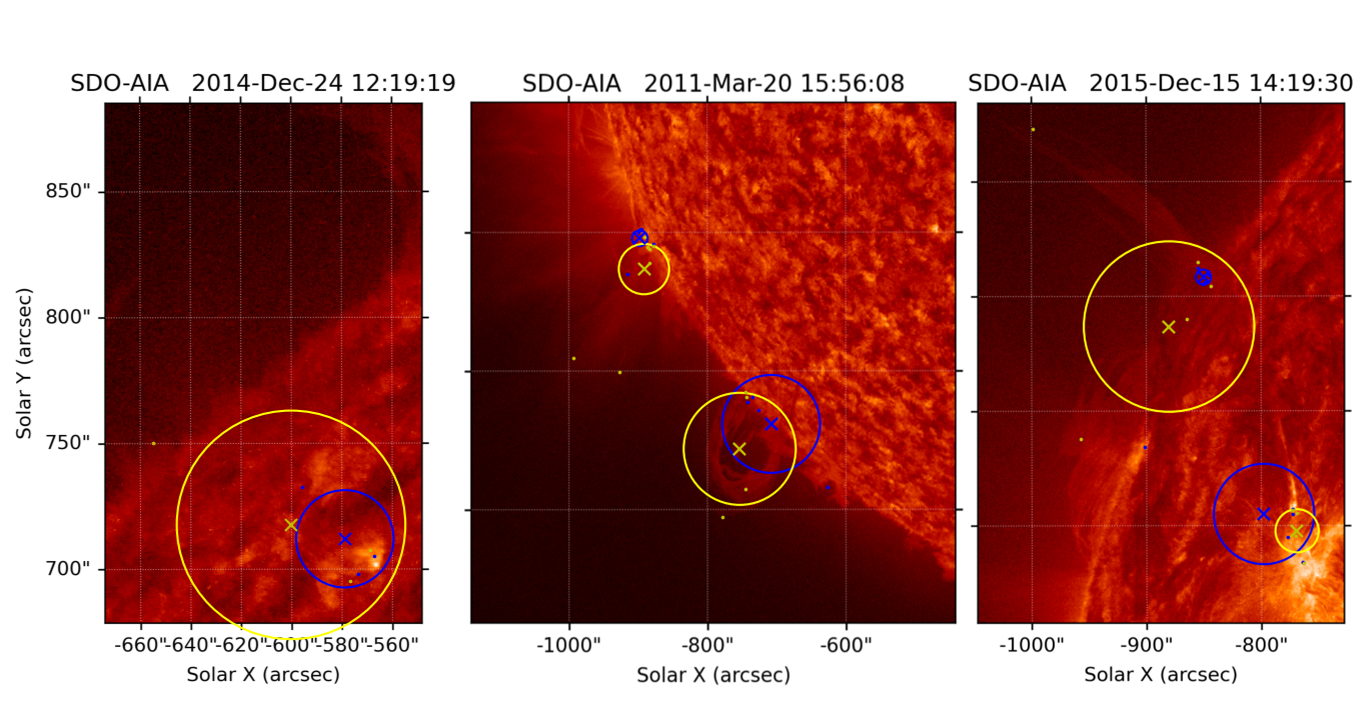}
    \caption{Examples of poorly performing start and end point clusters. These generally correspond to subjects where the jet is not clearly visible (or is actually not a jet).}
    \label{fig:worst_points}
\end{figure}

Figure~\ref{fig:worst_points} shows subjects with large point uncertainty ($\sigma_{\rm start} > 100$ pixels). These generally correspond to subjects where the jet is not clearly visible, or there is large confusion on whether the feature is even a jet. 

\subsection{Box clustering} \label{appendix:box_clustering}
\begin{figure}
    \centering
    \includegraphics[width=\columnwidth]{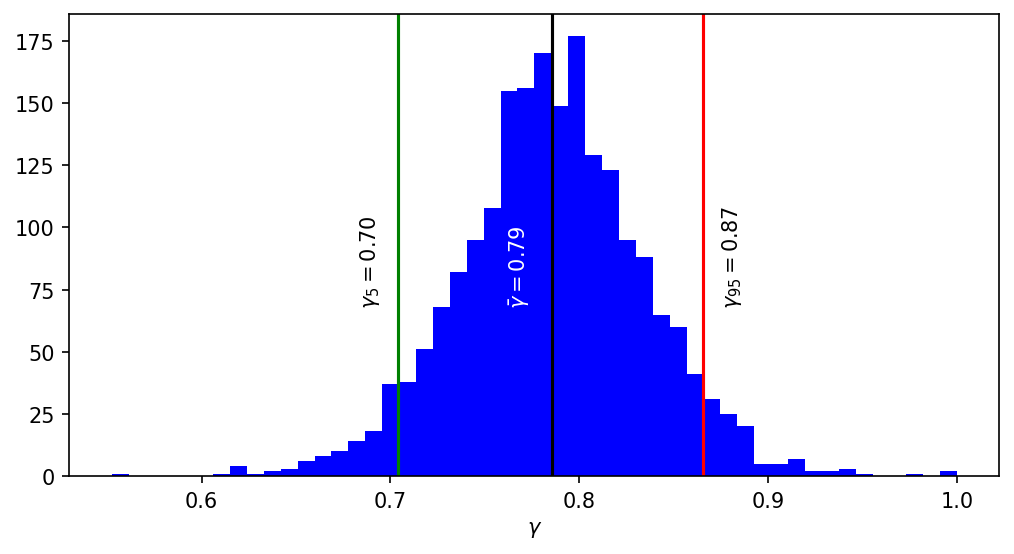}
    \caption{Distribution of $\gamma$ ($1\sigma$ uncertainty in the scale of the box) across all jets determined by our clustering algorithm.}
    \label{fig:gamma_iou}
\end{figure}

As described in \S~\ref{sec:box_clustering}, we use the Jaccard metric and DBSCAN algorithm to cluster the volunteer annotated boxes. The average box is determined by minimizing the sum Jaccard metric between the `average' box and all boxes in the cluster, and the corresponding uncertainty is given by Eq~\ref{equation:sigma_iou}. To provide physical meaning to $\sigma_{\rm IoU}$, we calculate $\gamma$ (the shape scale) as,
\begin{equation}
    \gamma = \sqrt{1 - \sigma_{\rm IoU}},
\end{equation}
where $\gamma$ corresponds to the 1$\sigma$ uncertainty in the scale of the average box (i.e., the largest and smallest box that fit within a 1$\sigma$ uncertainty are given by scaling the average box by a factor of $1 / \gamma$ and $\gamma$, respectively).
Figure~\ref{fig:gamma_iou} shows the distribution of $\gamma$ for all the jets annotated by vortices, along with the mean, 5th and 95th percentiles. The mean $\gamma$ is 0.8 (i.e., the $1\sigma$ bounds are $\sim20\%$ smaller and larger respectively). 

\begin{figure}
    \centering
    \includegraphics[width=\columnwidth]{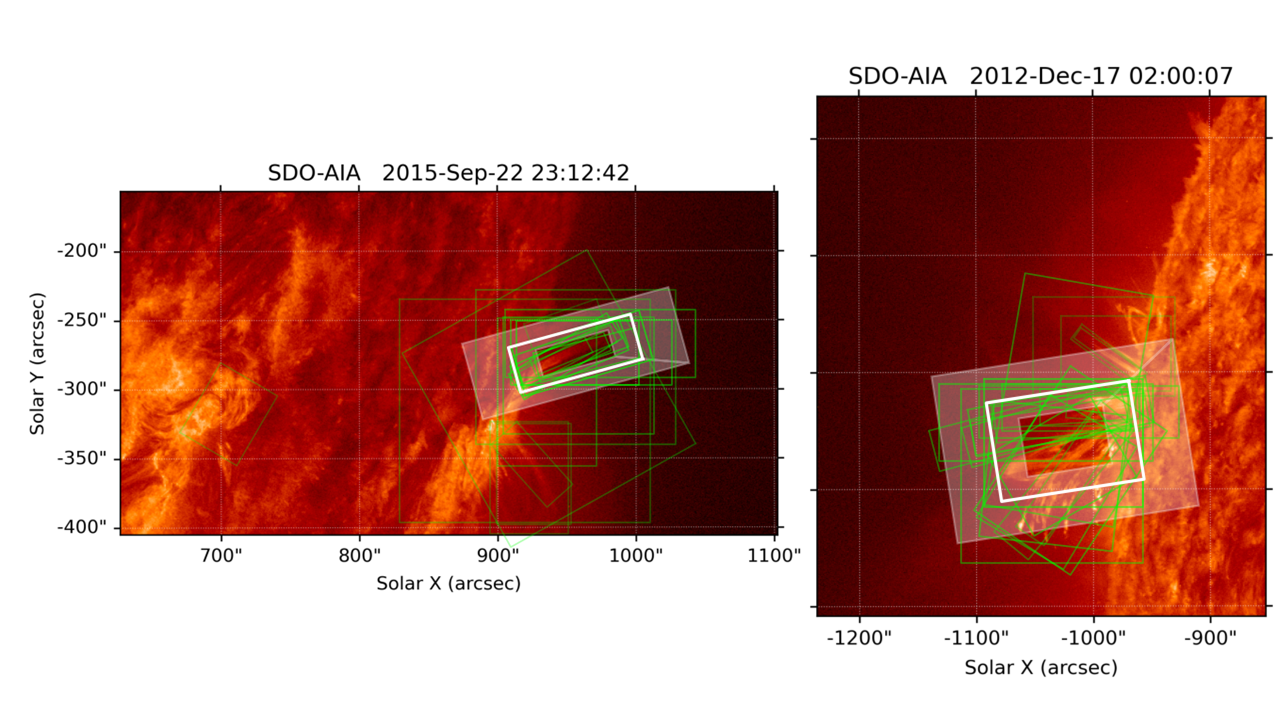}
    \caption{Examples of poor consensus on boxes. These generally correspond to locations where there are multiple overlapping jet-like features resulting in higher confusion for the volunteers on where the annotations should be.}
    \label{fig:box_cluster_examples}
\end{figure}

Most of the poorly performing subjects (low gamma) correspond similarly to subjects where the jet is not clearly visible or there are lots of competing regions annotated by volunteers. Two examples of low $\gamma$ subjects are shown in Figure~\ref{fig:box_cluster_examples}. 

\bibliographystyle{aa} 
\bibliography{zmabiblio} 

\end{document}